\documentclass[conference]{IEEEtran}

\usepackage{amsmath,amssymb,amsfonts}
\usepackage[linesnumbered,ruled,vlined,nofillcomment]{algorithm2e}
\usepackage{graphicx}
\usepackage{textcomp}
\usepackage{xcolor}
\usepackage{comment}
\usepackage{mathrsfs} 
\usepackage[numbers, square, comma, sort&compress]{natbib}
\usepackage{enumitem}

\DeclareMathOperator*{\argmax}{argmax}

\newcommand{\eccode}[1]{\mathcal{C}}
\newcommand{\codebook}[1]{\mathcal{C}}
\newcommand{\alphabet}[1]{\{0,1\}}
\newcommand{\codeword}[1]{\mathbf{C}}
\newcommand{\softinfo}[1]{\mathbf{R}}
\newcommand{\bestchase}[1]{\mathbf{D}}
\newcommand{\decodelist}[1]{\mathscr{L}}
\newcommand{\expected}[1]{\mathbb{E}[#1]}

\newcommand{\shannonEntropy}{H}
\newcommand{\oneornot}{z^{n}_{i}}
\newcommand{\listsize}{L}
\newcommand{\eBCH}{\text{eBCH}}
\newcommand{\BCH}{\text{BCH}}
\newcommand{\RLC}{\text{RLC}}
\newcommand{\CRC}{\text{CRC}}
\newcommand{\kvec}[1]{\mathbf{#1}}

\SetKw{KwReturn}{return}
\SetKw{KwBreak}{break}
\SetKw{KwYield}{yield}
\SetKw{KwOr}{or}
\SetKwComment{Comment}{// }{}

\SetCommentSty{mycommfont}

\graphicspath{{./img/}}

\def\BibTeX{{\rm B\kern-.05em{\sc i\kern-.025em b}\kern-.08em
    T\kern-.1667em\lower.7ex\hbox{E}\kern-.125emX}}

\begin{document}

\title{Block turbo decoding with ORBGRAND}

\author{
    \IEEEauthorblockN{Kevin Galligan}
    \IEEEauthorblockA{\textit{Hamilton Institute}\\
                    Maynooth University, Ireland\\
                    kevin.galligan.2020@mumail.ie}
    \and
    \IEEEauthorblockN{Muriel Médard}
    \IEEEauthorblockA{\textit{Research Laboratory of Electronics}\\
                    Massachusetts Institute of Technology, USA\\
                    medard@mit.edu}
    \and
    \IEEEauthorblockN{Ken R. Duffy}
    \IEEEauthorblockA{\textit{Hamilton Institute}\\
                    Maynooth University, Ireland\\
                    Ken.Duffy@mu.ie}
}

\maketitle

\begin{abstract}
Guessing Random Additive Noise Decoding (GRAND) is a family of universal decoding algorithms suitable for decoding any moderate redundancy code of any length. We establish that, through the use of list decoding, soft-input variants of GRAND can replace the Chase algorithm as the component decoder in the turbo decoding of product codes. In addition to being able to decode arbitrary product codes, rather than just those with dedicated hard-input component code decoders, results show that ORBGRAND achieves a coding gain of up to 0.7dB over the Chase algorithm with same list size.
\end{abstract}


\section{Introduction}
GRAND is a recently introduced family of hard \citep{duffy_capacity-achieving_2019,An20,galligan2021} and soft \cite{Duffy19a,solomon20,Duffy21_ordered,duffy2022_ordered} decoding algorithms that can accurately decode arbritrary codes, even non-linear ones \cite{cohen2022_aes}, with a moderate number of redundant bits. Their promise for practical, highly-parallelised decoding has led to the publication of several circuit designs and a chip implementation \citep{abbas2020grand,abbas2021orbgrand,abbas2021grandmo,riaz2021multicodegrand,condo2021fixed}.


Product codes are a class of long, high redundancy codes that are constructed by concatenating shorter component codes. Iterative GRAND (IGRAND) \cite{galligan2021} adapts GRAND for accurate hard-input iterative decoding of product codes \cite{elias_error-free_1954}. Here we establish how GRAND can decode product codes with the aid of soft information. Block turbo decoding achieves near-optimal soft-input decoding of product codes. Block turbo decoding uses a soft-input component decoder, customarily the Chase algorithm \cite{chase1972class}, to generate a selection of candidate decodings for a row or column of the product code, which are used to update the reliability of each bit in that row or column.




We reconsider soft-input GRAND as a list decoding algorithm and use it to replace the Chase algorithm. The decoding list produced by GRAND can be used to update the bit reliabilities as before, with the benefit that GRAND's code-agnosticism allows it to turbo decode any product code. We present results for block turbo decoding with Ordered Reliability Bits GRAND (ORBGRAND) \cite{Duffy21_ordered,duffy2022_ordered,abbas2021listgrand,condo2021_highperformance,condo2022_iterative}, a soft-input variant of GRAND that is particularly suited to hardware implementation \cite{abbas2021orbgrand,condo2021fixed}. We also provide analytical support for list decoding with GRAND algorithms and consider the standalone list decoding performance of ORBGRAND.

\section{Background}
\label{sec:turbo-background}
GRAND is a class of decoding algorithms for channel coding that concern themselves with the effects of noise. The core idea is to sequentially generate, from most to least likely, the binary noise effects that potentially corrupted the original message, based on knowledge of channel statistics or soft information. Each noise effect $z^n$ is removed from the demodulated channel output $y^n$, producing $\hat{x}^n=y^n\ominus z^n$. If $\hat{x}^n$ is in the codebook, then the noise effect $z^n$ is the most likely one to have occurred, and $\hat{x}^n$ is returned as a maximum-likelihood decoding. GRAND is code-agnostic because this guessing process does not depend on code structure, and requires only a method of checking codebook membership, such as a syndrome computation for linear codes \cite{lin_error_2004}.



Both hard-input and soft-input variants of GRAND exist. Of interest here is ORBGRAND, which works as follows. Given rank-ordered reliability values $\{r_i\}$ for the demodulated bits $y^n$, where $r_i\ge 0$ and $r_i<r_j$ for $i<j$, the likelihood of a putative noise sequence $z^n$ is proportional to $f(z^n)=\sum_{i=1}^n r_i \oneornot{}$. The basic version of ORBGRAND approximates the reliability values with a line through the origin, $r_i=\beta i$, $\beta \ge 0$, and based on this model it efficiently generates noise sequences in approximate maximum-likelihood order. The full version of the algorithm instead uses a multiline model. Here, we consider the 1-line model, which distinguishes itself from basic ORBGRAND by potentially having a non-zero intercept. 1-line ORBGRAND uses the approximation $r_i = \alpha + \beta i$, where $\alpha,\beta \ge 0$. Then $f(z^n)=\sum_{i=1}^n (\alpha + \beta i)\oneornot{} = \alpha\sum_{i=1}^n\oneornot{} + \beta\sum_{i=1}^n i\oneornot{} = \alpha w_H(z^n) + \beta w_L(z^n)$, where $w_H(z^n)$ is the Hamming weight and $w_L(z^n)$ is the \textit{logistic weight} of $z^n$. We assume $c=\alpha/\beta$ to be a non-negative integer, and let $w_T(z^n) = f(z^n)/\beta = c w_H(z^n) + w_L(z^n)$ be the \textit{total weight} of a noise sequence, where $w_T(z^n) \in \{0,1,2,...\}$.

The full ORBGRAND algorithm \cite{duffy2022_ordered} details how to efficiently generate putative noise sequences in order of their total weight, which is an approximation of their maximum-likelihood order, without the need for dynamic memory. We detail a simple method for the 1-line version in section \ref{sec:1line-for-turbo}, and also a method to find a suitable value for $c$.

A 2-dimensional product code \cite{elias_error-free_1954} is a concatenation of two systematic component codes, $\eccode{}_i$, for $1\le i \le 2$. $\eccode{}_i$ has parameters $[n_i, k_i, R_i, d_i]$, where $n_i$ is its code length, $k_i$ is its number of information symbols, $R_i=k_i/n_i$ is its code rate, and $d_i$ is its minimum Hamming distance. The input symbols are arranged as a $k_2 \times k_1$ array. The rows of this array are extended by encoding them with $\eccode{}_1$, then the columns are extended by encoding them with $\eccode{}_2$. The result is an $n_2 \times n_1$ array, all rows and columns of which are codewords of $\eccode{}_1$ and $\eccode{}_2$. This is a codeword of the product code, with parameters $[n_1n_2, k_1k_2, R_1R_2, d_1d_2]$. We use $C(n,k,d)^2$ to denote a product code with row and column code $C$, where $C$ has parameters $[n,k,k/n,d]$. Pyndiah \cite{pyndiah_1998} extended the turbo decoding technique from convolutional codes to product codes, where row decoding output informs the soft input of the column decoding, and vice versa, as we describe later. Hard-input GRAND algorithms have been applied to product codes and related coding structures \cite{galligan2021,chatzigeorgiou2022_transversal,allahkaram2022_urllc}.

\section{Turbo and List Decoding With GRAND}
\label{sec:1line-for-turbo}



List decoding \cite{elias1957_list} is a decoding procedure in which the decoder outputs a list of $L$ codewords. The original GRAND algorithm stops the guessing process once it finds a codeword, since that codeword is a maximum-likelihood decoding. In a form of list decoding, Abbas et al. \cite{abbas2021listgrand} recently proposed an extension to basic ORBGRAND in which codewords are accumulated within some distance of the first codeword that is found and the resulting list is used to improve hard-output accuracy. Here, we provide analytical support for list decoding with any GRAND algorithm, based on theorems from \cite{duffy_capacity-achieving_2019}. GRAND can list decode by continuing its guessing process until it has accumulated $L$ codewords, rather than stopping after the first one, described in Algorithm \ref{algo:list-grand}.


\begin{algorithm}
	\caption{GRAND list decoding of hard channel output $y$, possibly with soft output $r$ informing the likelihood of noise effects. Given a codebook $\codebook{}$, code length $n$ and a target list size $\listsize{}$.}
	\label{algo:list-grand}
	$\decodelist{} \gets \{\}$\;
	$z^* \gets 0^n$\Comment*[l]{all-zero is most likely}
	\While{$\lvert \decodelist{} \rvert < L$}{
		$c^* \gets y \ominus z^*$\Comment*[l]{undo noise effect}
		\If{$c^* \in \mathcal{C}$}{
			add $c^*$ to $\decodelist{}$\;
		}
		$z^* \gets$ next most likely noise effect\;
	}
	\KwReturn $\decodelist{}$\;
\end{algorithm}
\vspace{-10pt}

Underlying GRAND is a race between two processes: the number of guesses to find the true channel noise effect $Z^n : \Omega \to \alphabet{}^n$, which recovers the correct codeword, and the number of guesses $U : \Omega \to \{1,...,2^n\}$ before GRAND identifies an incorrect codeword. GRAND's guessing order is defined by a bijective function $G:\alphabet{}^n \to \{1,...,2^n\}$ that maps each noise effect to its position in the guessing order. The number of guesses to identify the true channel noise effect is $G(Z^n)$. GRAND identifies the correct decoding when $G(Z^n) < U$, the asymptotic probability of which as $n$ tends to infinity is derived in \cite{duffy_capacity-achieving_2019} for uniform random codebooks.


We now examine the case of list decoding with GRAND, and its approximate complexity. Consider a random binary code of length $n$ with $2^k$ codewords. For list size $L=2^l$, denote the position of the $i$-th incorrect codeword in GRAND's guessing order by the random variable $U_i : \Omega \to \{1, ..., 2^n\}$, where $1\le i \le 2^k-1$. As the codebook is uniformly at random, the $\{U_i\}$ appear uniformly in the guesswork order $\{1,...,2^n\}$. Let the codewords be ordered such that $U_1 < U_2 < ... < U_{2^k-1}$. Given hard channel output $Y^n : \Omega \to \{0,1\}^n$ and the $i$-th codeword $\codeword{}_i \in \codebook{}$, $U_i = G(Y^n \oplus \codeword{}_i)$, since $\codeword{}_i=Y^n\ominus(Y^n+\codeword{}_i)$. The total number of guesses to accumulate $L$ codewords is $\Upsilon_L = U_1 + \sum_{i=2}^L (U_i - U_{i-1})$.



The expected total number of guesses $\expected{\Upsilon_L}$ is derived as follows. Since $\expected{U_1} = \sum_u \expected{U_1 | U_2=u} P(U_2=u) = (1/2)\sum_u u P(U_2=u) = \expected{U_2}/2$, the expected guesses from the first codeword to the second is $\expected{U_2-U_1}=\expected{U_2}-\expected{U_1}=\expected{U_2}/2 = \expected{U_1}$. A similar argument proves $\expected{U_{i}-U_{i-1}}=\expected{U_1}$ for all $i$, and $\expected{2^n-U_{2^k-1}}=\expected{U_1}$, which is the expected number of guesses from the final codeword to the last binary sequence that GRAND can guess; thus, $\expected{\Upsilon_L}=L\expected{U_1}$. The expected number of guesses to cover all $2^n$ possible noise effects is $\expected{\sum_{i=1}^{2^k}U_i-U_{i-1}} = 2^k \expected{U_1} = 2^n$, where $U_0=0$ and $U_{2^k}=2^n$. Hence $\expected{U_1} = 2^{n-k}$ and $\expected{\Upsilon_L} = L \expected{U_1} = 2^{n-k+l}$.

The above argument informs the choice of list size and code rate in coding scheme design. As $n$ becomes large, $G(Z^n) \leq 2^{n\shannonEntropy}$ with high likelihood, where $\shannonEntropy$ is the Shannon entropy of the channel noise \cite{christiansen2013_guesswork}. The correct codeword ends up on the decoding list with high likelihood when $G(Z^n) < \expected{\Upsilon_L}$, which is true when $2^{nH} < \expected{\Upsilon_L} = 2^{n-k+l} = 2^{n(1-R)+l}$. Letting $l=n\theta$ for $\theta>0$, the requirement becomes $2^{nH} < 2^{n(1-R+\theta)}$, or $H < 1 - R + \theta$. Stated in a form closer to the noisy-channel coding theorem \cite{shannon1948}, $R < 1-H+\theta$. This tells us that by increasing the list size we can perform effective channel coding at higher code rates, as asserted in \cite{elias1957_list}.

Regarding complexity, $2^{n-k+l}$ is an upper bound on the expected number of GRAND queries for random codebooks. From this arises a design trade-off between list size and the number of parity bits. To keep the bound constant, a parity bit must be removed if the list size is doubled. The bound corresponds to the expected number of queries to identify $L$ incorrect codewords, although in practice the correct codeword will typically be added to the list after a small number of queries and fewer overall queries will be required as a result. 

We now turn to the block turbo decoding algorithm of Pyndiah \cite{pyndiah_1998}. The decoding of a single component (row or column) of the product code, with accompanying per-bit soft information $\softinfo{} \in \mathbb{R}^n$, works as follows:

\begin{enumerate}[noitemsep,topsep=0pt,parsep=0pt,partopsep=0pt,leftmargin=*]
    \item Apply Chase decoding to produce a list $\decodelist{} \subseteq \codebook{}$, where 
    $1 \leq \lvert \decodelist{} \rvert \leq 2^\rho$ for some small positive integer $\rho$.
    \item Select $\bestchase{} = \argmax_{\codeword{} \in \decodelist{}} \lvert \softinfo{} - \codeword{} \rvert^2$ as the new value of the component, where $\lvert \softinfo{} - \codeword{} \rvert^2$ is the Euclidean distance between $\softinfo{}$ and the modulated form of $\codeword{}$.
    \item Individually update the soft information of each bit in the component. If there is a codeword that disagrees with $\bestchase{}$ on the value of the $i$-th bit, $\codeword{}^* \in \argmax_{\{\codeword{} \in \decodelist{} : \codeword{}_i \neq \bestchase{}_i\}} \lvert \softinfo{} - \codeword{} \rvert^2$, then the soft output for that bit is given by $r_i = \bestchase{}_i (\lvert \softinfo{} - \codeword{}^* \rvert^2 - \lvert \softinfo{} - \bestchase{} \rvert^2)/4$.
    \item If $\argmax_{\{\codeword{} \in \decodelist{} : \codeword{}_i \neq \bestchase{}_i\}} \lvert \softinfo{} - \codeword{} \rvert^2$ is empty, then instead use $r_i = \beta$ for some constant $\beta \geq 0$.
\end{enumerate}

A soft-input list decoding variant of GRAND can replace the Chase algorithm in step (1) of the block turbo decoding algorithm, 
with the remainder of the algorithm untouched. Indeed, a variant of ORBGRAND has independently been proposed \cite{condo2022_iterative} for use in Pyndiah-style soft-input soft-output iterative decoding of OFEC codes, another form of concatenated code. Iterative soft-input soft-output GRAND decoding has also recently been considered in \cite{sarideddeen2022_softinput}.

An advantage of GRAND as a component decoder is that it can decode any component code, and thus can turbo decode any product code. The Chase algorithm requires that a specialised hard-input decoder exists for the component codes, which for example is not the case for Random Linear Codes (RLCs) and CRC codes. GRAND also distinguishes itself by populating its list with codewords in maximum-likelihood order, assuming its query order is correct, while Chase makes no such guarantees and may output duplicate codewords.

\begin{figure}
	\centering
	\includegraphics[scale=0.6]{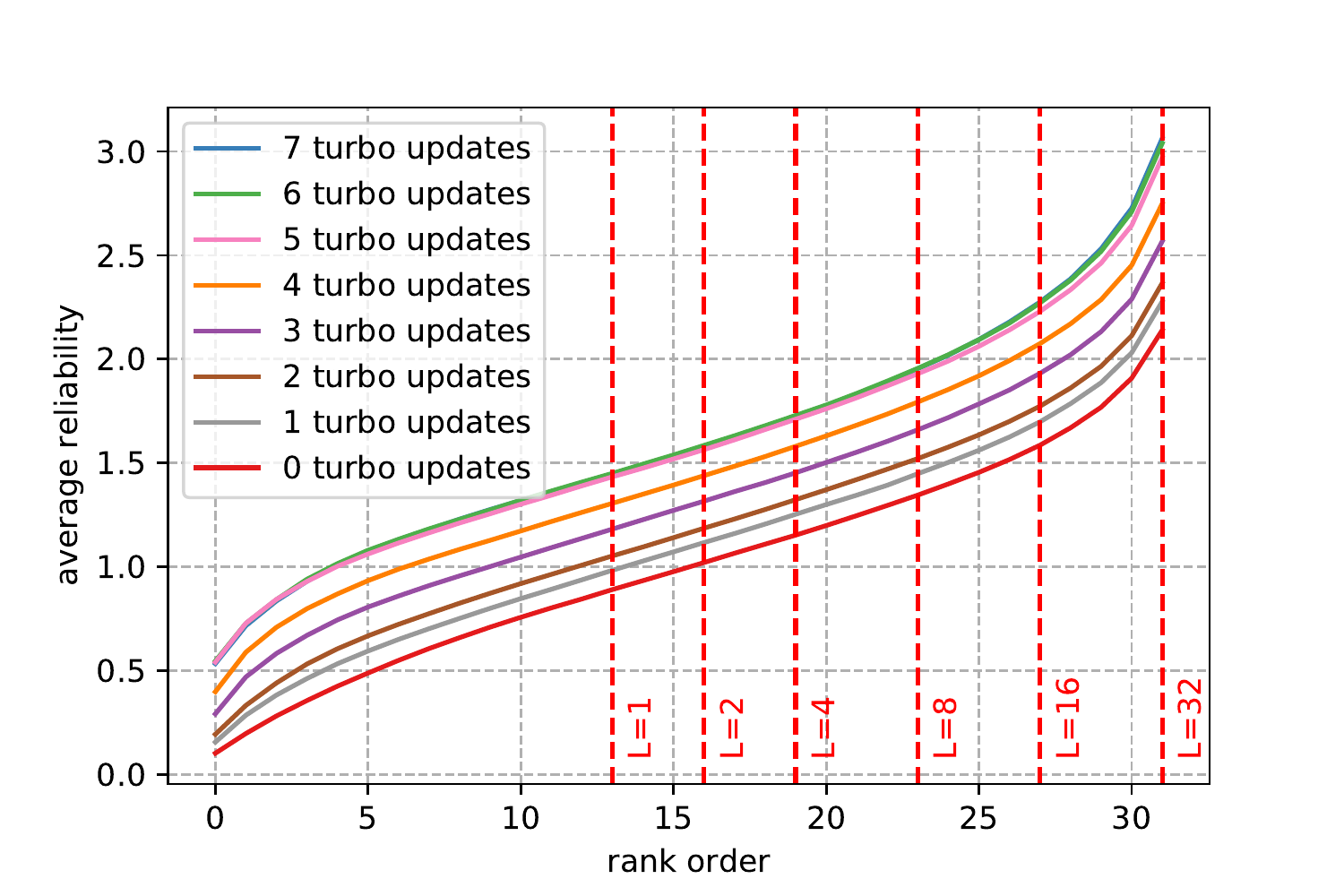}
	\caption[Distribution of soft input during turbo decoding.]{The average distribution of rank-ordered soft input throughout turbo decoding of an $\eBCH(32,26,4)^2$ product code with a Chase component decoder, where $E_b/N_0=4$dB and $\rho=4$. The distribution converges around the 5th update. Dotted red lines mark the max-ranked bit that, on average, ORBGRAND would be expected to flip, given list size $L$.}
	\label{fig:llr-distribution}
\end{figure}

ORBGRAND is a practical component decoder for turbo decoding, given its accuracy and efficiency. As turbo decoding converges, however, the distribution of reliability values shifts upwards, making basic ORBGRAND's linear approximation less suitable, as in Fig. \ref{fig:llr-distribution}. We thus propose to perform turbo decoding with the full ORBGRAND algorithm, parameterised to use a single line. This enables ORBGRAND's model to have a non-zero intercept and so better capture the input reliability distribution during later iterations of turbo decoding.

In Algorithm \ref{algo:shift-orbgrand}, we present a simple method to construct noise sequences for 1-line ORBGRAND that has similar implementation complexity as basic ORBGRAND. Noise sequences are generated in order of their total weight, $w_T$. For each $w_T$, we iterate over all pairs of non-negative integers $(w_H, w_L)$ such that $w_T = c w_H + w_L$, and the landslide algorithm \cite{duffy2022_ordered} generates all noise sequences for each pair.

\begin{algorithm}
	\caption{Noise effect generation algorithm for 1-line ORBGRAND, given integer parameter $c\ge 0$ and code length $n$.}
	\label{algo:shift-orbgrand}
	\KwYield $0^n$\Comment*[l]{all-zero is most likely}
	$w_T \gets c + 1$\Comment*[l]{minimum possible weight}
	\While{$w_T \le cn + \frac{n(n+1)}{2}$}{
		$w_H \gets \max(1,\lceil\frac{1+2(n+c)-\sqrt{(1+2(n+c))^2 - 8w_T}}{2}\rceil)$\;
		\While{$w_H \le n$}{
			$w_L \gets w_T - c w_H$\;
			\If{$w_L \le 0$ \KwOr $w_L < \frac{w_H(w_H+1)}{2}$}{
				\KwBreak\Comment*[l]{invalid pair}
			}
			\KwYield noise effects generated by $\texttt{Landslide}(w_H, w_L, n)$\;
			$w_H \gets w_H + 1$\;
		}
		$w_T \gets w_T + 1$\;
	}
\end{algorithm}

A simple and effective method to pick $c$ is to fit a line through the points $(1,r_1)$ and $(\lfloor n/2 \rfloor, r_{\lfloor n/2 \rfloor})$, where $n$ is the code length. Then $\beta=(r_{\lfloor n/2 \rfloor}-r_1)/(\lfloor n/2 \rfloor -1)$ and $c=\max(0, [(r_1-\beta)/\beta])$, where $[.]$ rounds to the nearest integer. This gives the best estimate of $(1,r_1)$, the least reliable and most important bit, and accurately approximates the remaining values if they follow a line-like distribution as in Fig. \ref{fig:llr-distribution}.



\section{Performance Evaluation}
\begin{figure}
\vspace{-0.2in}
	\centering
	\includegraphics[scale=0.6]{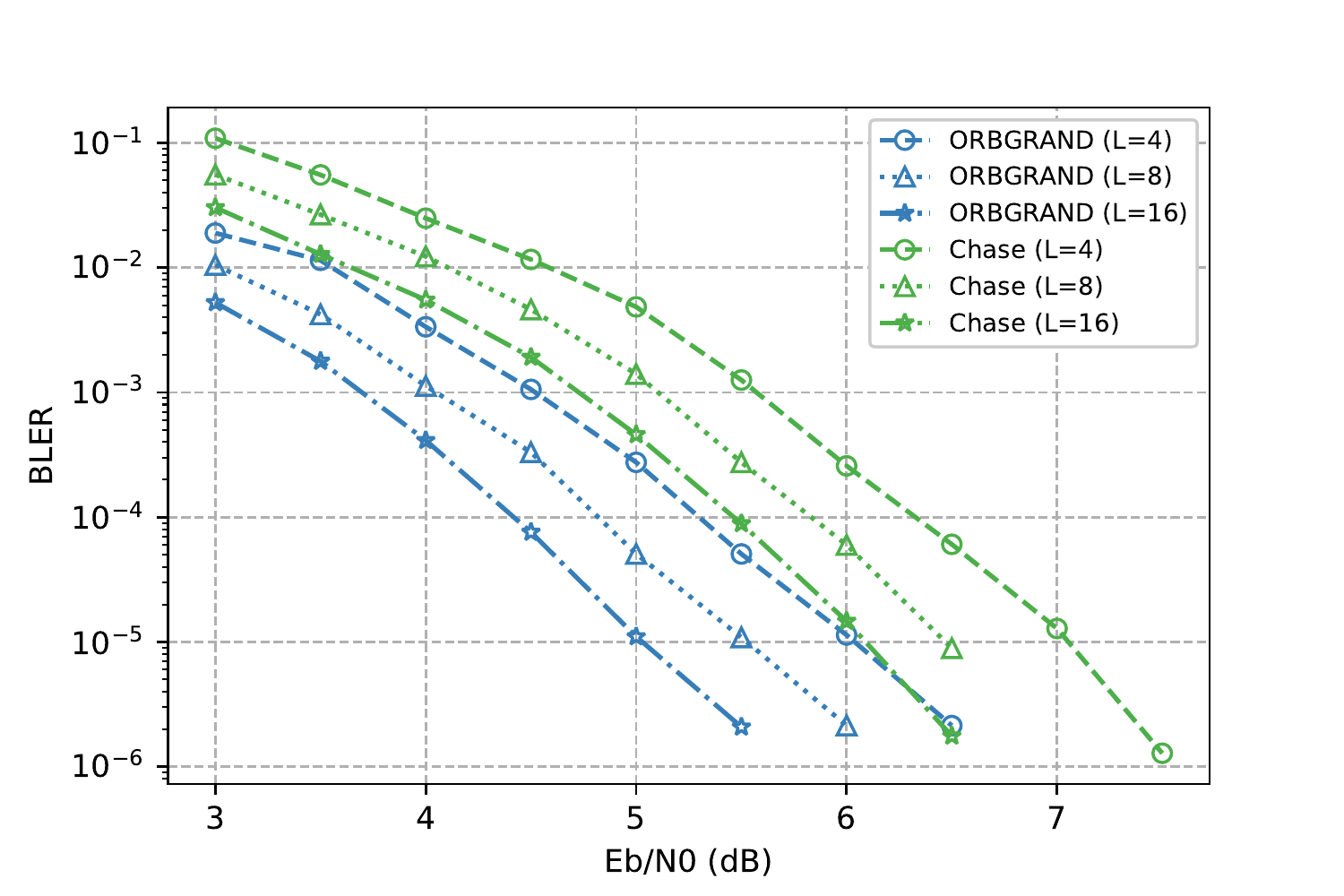}
	\caption[List decoding BLER of an $\eBCH(32,26,4)$ code with basic ORBGRAND and Chase as list decoders.]{List decoding BLER of an $\eBCH(32,26,4)$ code with ORBGRAND and Chase as list decoders, and list size $L$. ORBGRAND consistently provides a coding gain over Chase, even with smaller list size.}
	\label{fig:list-orbgrand-bler}
\vspace{-0.1in}
\end{figure}

We begin with a study of ORBGRAND's list decoding performance, since this is critical to its performance as a turbo component decoder. We run simulations in an additive white Gaussian noise channel with binary phase shift keying modulation. Fig. \ref{fig:list-orbgrand-bler} shows the list decoding block error rate (BLER) of basic ORBGRAND versus that of Chase decoding for an extended BCH code  \cite{lin_error_2004}, $\eBCH(32,26,4)$, which is one of the component codes from \cite{pyndiah_1998}. A list decoding block error occurs when, given transmitted codeword $\kvec{C}$ and output decoding list $\decodelist{}$, $\kvec{C} \notin \decodelist{}$. The list size $L$ ranges from 4 to 16 and $L=2^l$ corresponds to a Chase parameter of $\rho=l$. At a BLER of $10^{-5}$ and $L=16$, basic ORBGRAND provides a coding gain of 1 dB over Chase. That ORBGRAND outperforms Chase as a list decoder is a promising indicator that it will be a good turbo component decoder.

\begin{figure}
\vspace{-0.1in}
	\centering
	\includegraphics[scale=0.6]{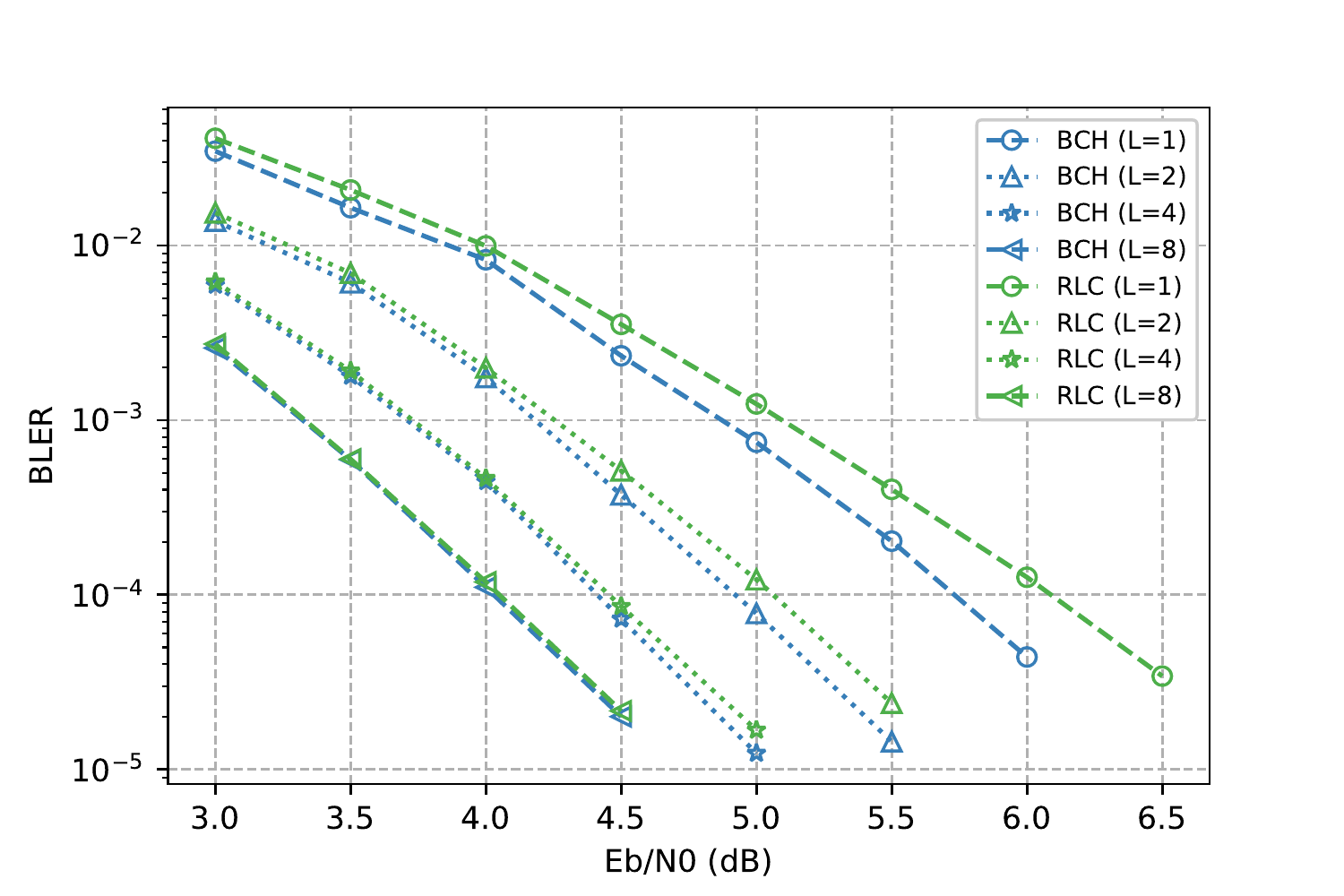}
	\caption[List decoding BLER of BCH and RLC codes.]{List decoding BLER of a $\BCH(31,21,5)$ code and an $\RLC(31,21,4)$ code, with basic ORBGRAND decoding and list size $L$. The RLC was purposely constructed to have a lower minimum distance, demonstrating that its performance converges to that of the BCH regardless of minimum distance.}
	\label{fig:bch-vs-rlc-list-decoding-bler}
\end{figure}

Fig. \ref{fig:bch-vs-rlc-list-decoding-bler} shows the basic ORBGRAND list decoding BLER of a BCH code, $\BCH(31,21,5)$, versus a random linear code, $\RLC(31,21,4)$. The RLC was purposely constructed to have a lower minimum distance. Its decoding is only enabled by ORBGRAND's code-agnosticism. As the list size increases, the performance of the two codes converges, which is consistent with Elias's claim that a larger list size should compensate for structural weakness in a code \cite{elias1957_list}. This suggests that powerful product codes can be constructed from imperfect component codes, but later results will demonstrate this is not true.

\begin{figure}
\vspace{-0.5cm}
	\centering
	\includegraphics[scale=0.6]{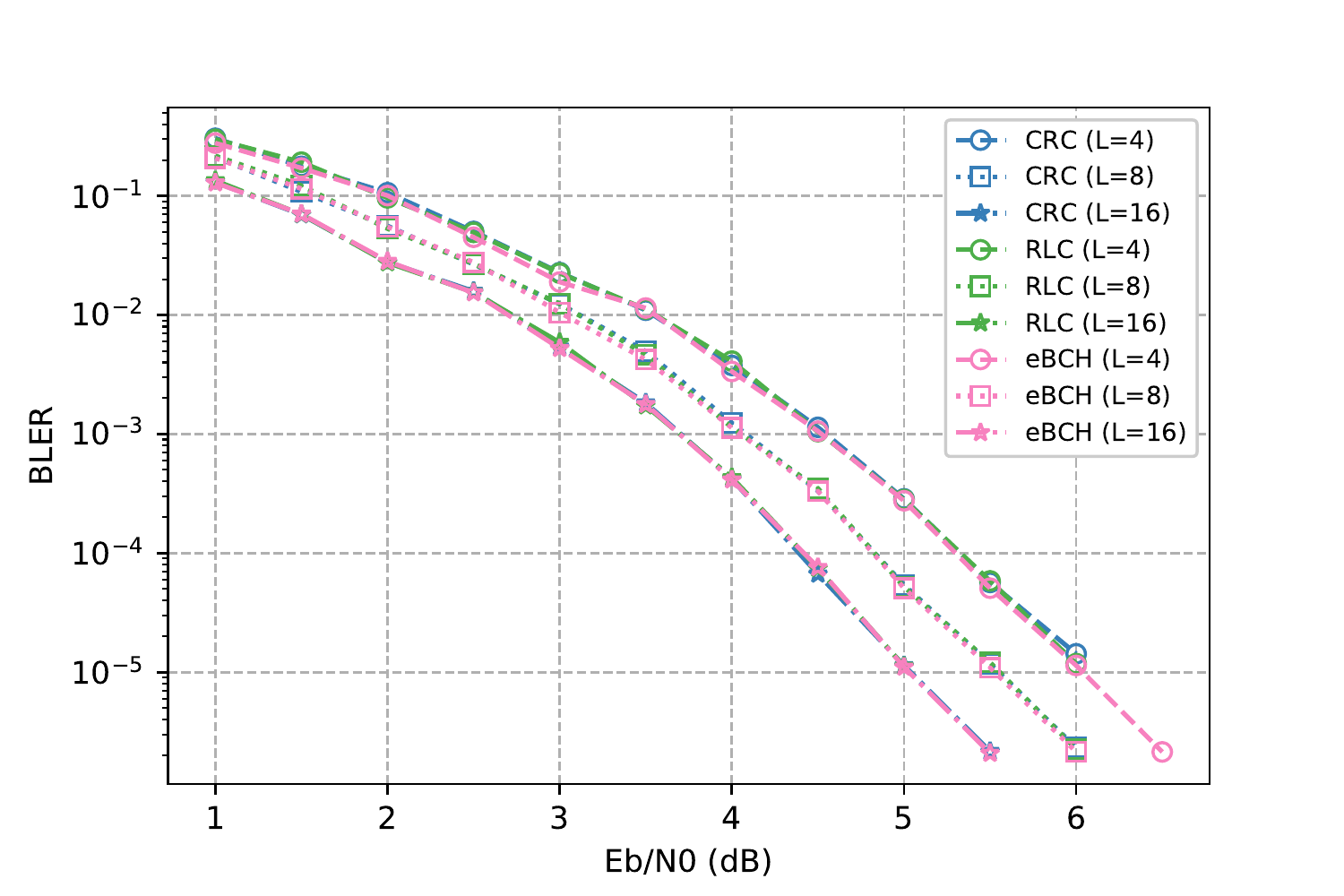}
	\caption[List decoding BLER of a selection of codes.]{List decoding BLER of $\eBCH(32,26,4)$, $\RLC(32,26,3)$ and $\CRC(32,26,3)$ codes, with basic ORBGRAND decoding.}
	\label{fig:list-bler-various-codes}
\vspace{-0.2cm}
\end{figure}

Fig. \ref{fig:list-bler-various-codes} shows the list decoding accuracy of a further selection of codes with basic ORBGRAND: $\eBCH(32,26,4)$, $\RLC(32,26,3)$ and $\CRC(32,26,3)$. The CRC polynomial is 0x33 in Koopman notation \cite{koopman2009cyclic}. The list size ranges from 4 to 16. Performance is essentially equivalent at these list sizes, despite the RLC and CRC having lower minimum distance.

\begin{figure}
\vspace{-0.1in}
	\centering
	\includegraphics[scale=0.6]{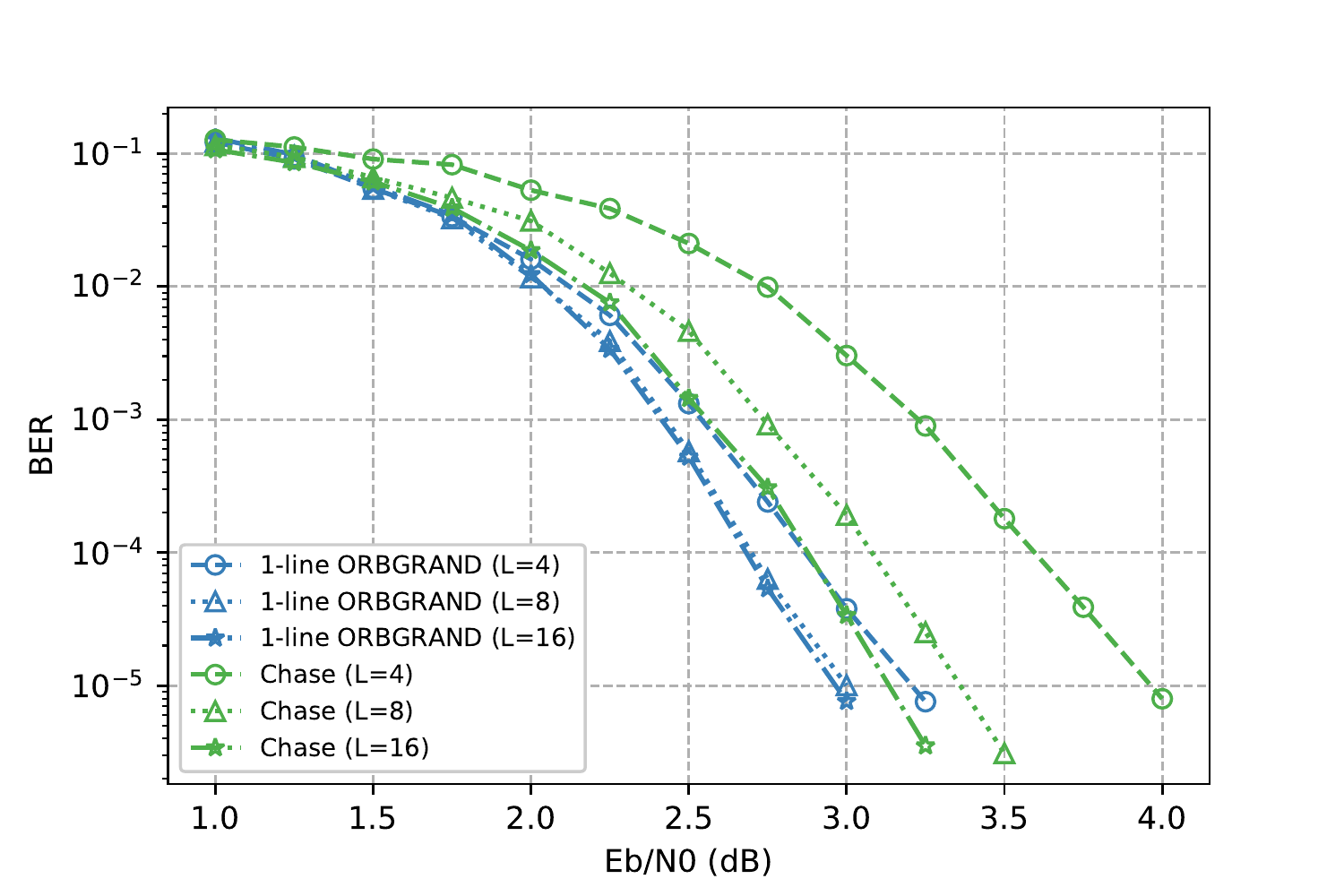}
	\caption[Block turbo decoding BER of an $\eBCH(32,26,4)^2$ product code.]{BER of an $\eBCH(32,26,4)^2$ product code ($n=1024, k=676, d=16$) with block turbo decoding. 1-line ORBGRAND and Chase are used as component decoders with list size $L$.}
	\label{fig:turbo-ber-ebch-32n}
\vspace{-0.1in}
\end{figure}

\begin{figure}
\vspace{-0.1in}
	\centering
	\includegraphics[scale=0.6]{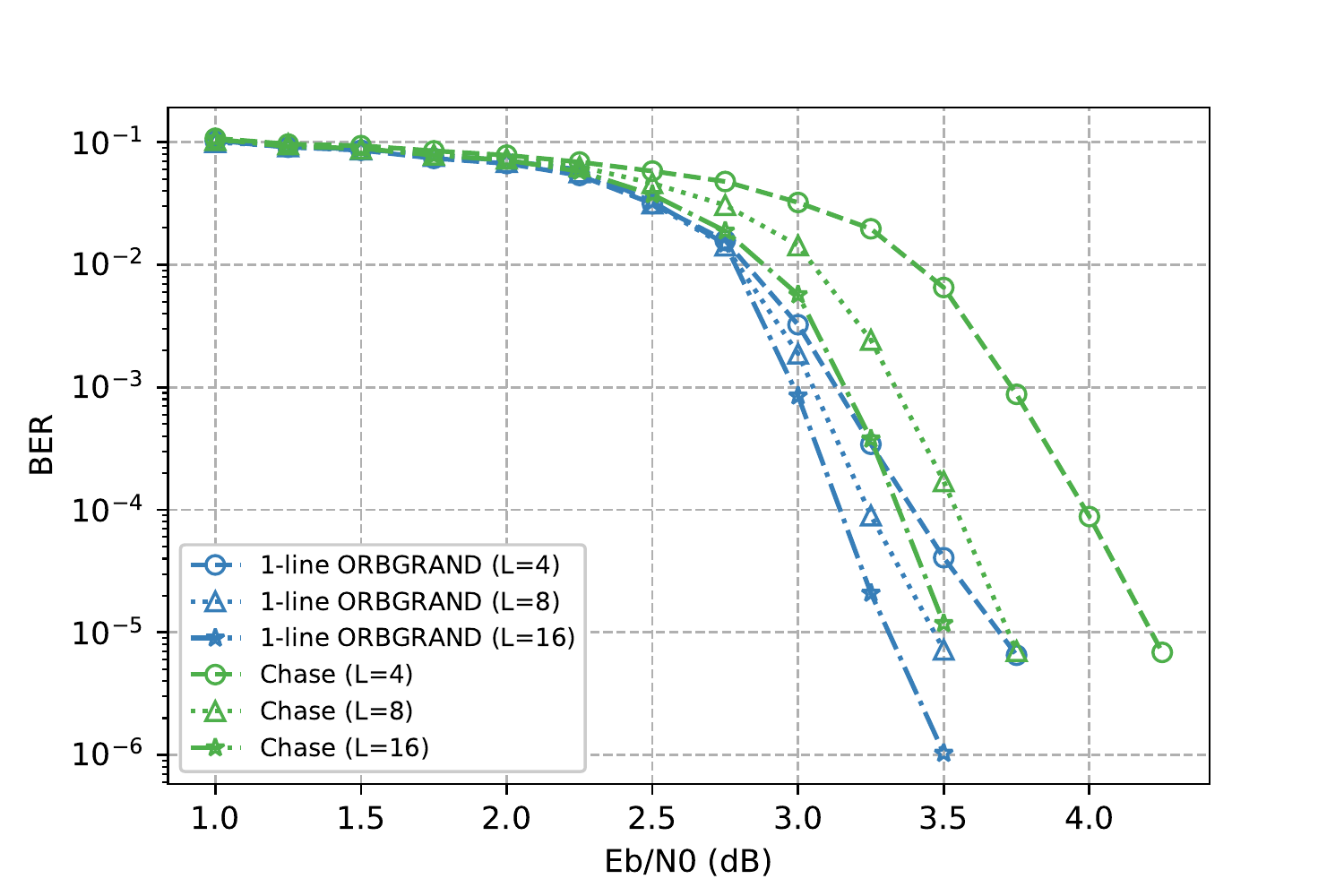}
	\caption[Block turbo decoding BER of an $\eBCH(64,57,4)^2$ code.]{BER of an $\eBCH(64,57,4)^2$ product code ($n=4096, k=3249,d=16$) with block turbo decoding. Decoding parameters are the same as those described in Fig. \ref{fig:turbo-ber-ebch-32n}. With $L=4$ the gain is 0.7 dB for BLER $10^{-4}$.}
	\label{fig:turbo-ber-ebch-64n}
\vspace{-0.1in}
\end{figure}

Fig. \ref{fig:turbo-ber-ebch-32n} and Fig. \ref{fig:turbo-ber-ebch-64n} show the bit error rate (BER) of $\eBCH(32,26,4)^2$ and $\eBCH(64,57,4)^2$ product codes with block turbo decoding. 1-line ORBGRAND and Chase are used as component decoders. These codes were tested in \cite{pyndiah_1998}, and, as in that paper, we run turbo decoding for 4 iterations. The list sizes are 4, 8 and 16. We utilise an early stopping criterion, so that decoding is halted if the rows or columns are found to be error-free at the beginning of any iteration.

Fig. \ref{fig:turbo-ber-ebch-32n} concerns an $\eBCH(32,26,4)^2$ code ($n=1024$, $k=676$, $R=0.66$). At $L=16$, 1-line ORBGRAND provides a coding gain over Chase of approximately 0.15dB at a BER of $10^{-5}$. Even with $L=4$, 1-line ORBGRAND achieves nearly the same performance, whereas the performance of Chase degrades rapidly as the list size decreases.

Fig. \ref{fig:turbo-ber-ebch-64n} shows results for an $\eBCH(64,57,4)^2$ code ($n=4096$, $k=3249$ $R=0.79$). Again, when $L=16$, 1-line ORBGRAND provides a coding gain of approximately 0.2dB at a BER of $10^{-5}$, and Chase performance degrades more severely with decreasing list size.

\begin{figure}
\vspace{-0.1in}
	\centering
	\includegraphics[scale=0.6]{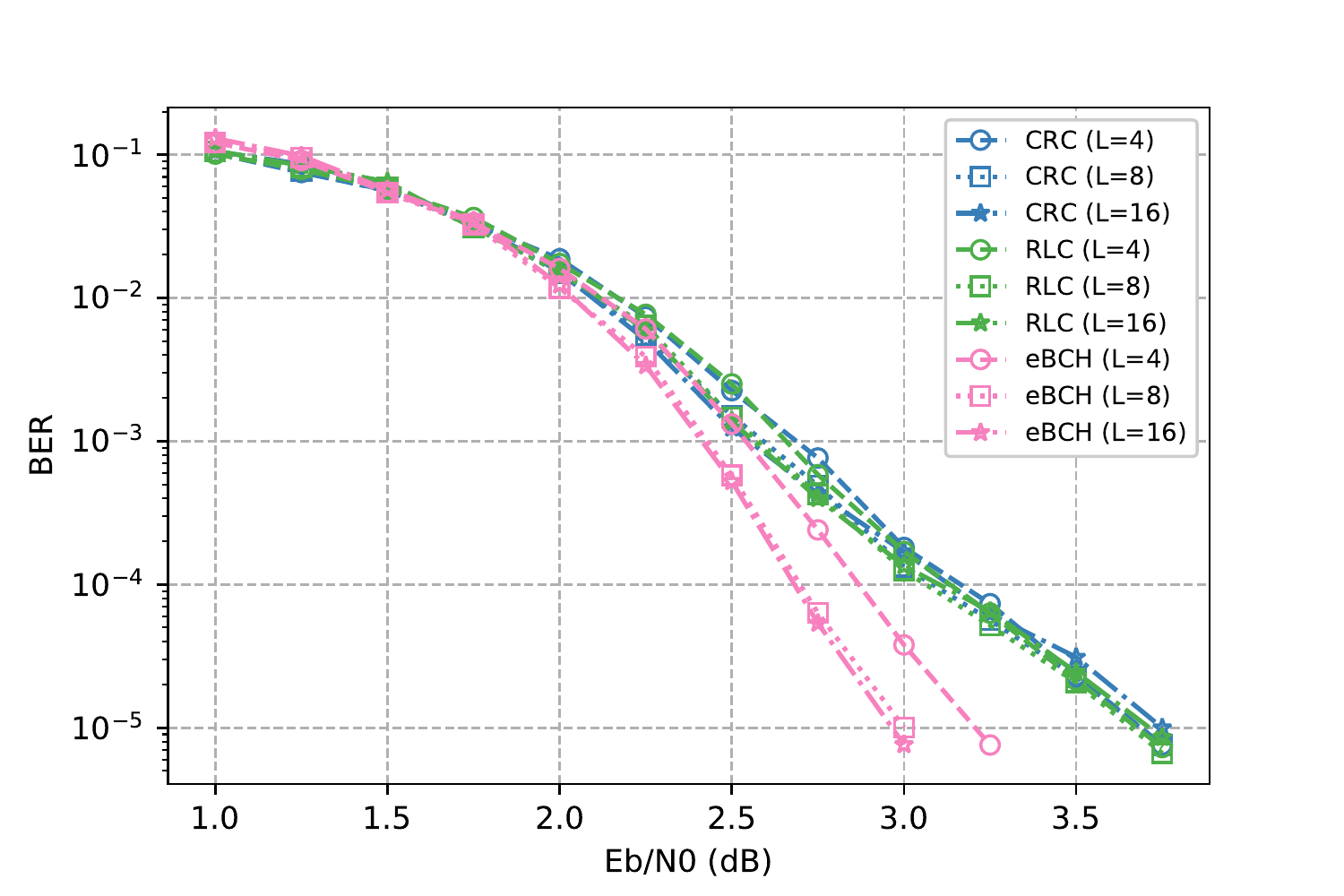}
	\caption[Block turbo decoding BER of a selection of product codes.]{BER of $\CRC(32,26,3)^2$, $\RLC(32,26,3)^2$, and $\eBCH(32,26,4)^2$ product codes with 1-line ORBGRAND turbo decoding, list size $L$.}
	\label{fig:turbo-ber-various-codes}
\vspace{-0.1in}
\end{figure}

Fig. \ref{fig:turbo-ber-various-codes} shows the turbo decoding accuracy of product codes whose component codes are the same as in Fig. \ref{fig:list-bler-various-codes}, with 1-line ORBGRAND decoding. Despite having equivalent list decoding performance, the RLC and CRC fare worse than the eBCH code as component codes, and so product codes appear to compound structural weakness in their component codes.

\section{Conclusions}
GRAND algorithms can list decode accurately, and soft-input list decoding GRAND algorithms are a viable replacement for Chase as the component decoder in block turbo decoding. We have presented a code for which basic ORBGRAND list decoding gains as much as 1dB over Chase at a BLER of $10^{-5}$. For turbo decoding, the distribution of rank-ordered soft information shifts so that the full version of ORBGRAND is required for effective component decoding. Turbo decoding simulations show that 1-line ORBGRAND gains up to 0.7dB over Chase at a BER of $10^{-5}$ for two different product codes. GRAND's universality allows list and turbo decoding to be applied to codes without bespoke decoders, such as CRCs and RLCs, as well as product codes that are concatenations of those codes. This leads to the possibility of new channel coding applications.

Acknowledgment: This publication has emanated from research conducted with the financial support of Science Foundation Ireland under grant number 18/CRT/6049. The opinions, findings and conclusions or recommendations expressed in this material are those of the author(s) and do not necessarily reflect the views of the Science Foundation Ireland. This work was partially supported by Defense Advanced Research Projects Agency contract number HR00112120008.


\bibliographystyle{IEEEtran}
{\small \bibliography{grand-list-decoding}}

\end{document}